\documentclass[runningheads]{llncs}
\usepackage[T1]{fontenc}
\usepackage{tgbonum}
\usepackage{tgcursor}
\usepackage{hyperref}

\usepackage{graphicx}
\usepackage{tikz,lipsum,lmodern}
\usepackage{array}
\usepackage{tabularx}
\usepackage[many]{tcolorbox} 
\newtcolorbox{boxK}{
    sharpish corners, 
    boxrule = 0pt,
    toprule = 4.5pt, 
    enhanced,
    fuzzy shadow = {0pt}{-2pt}{-0.5pt}{0.5pt}{black!35} 
}
\begin{document}
\title{LLMs' ways of seeing User Personas}
%
%
\author{Swaroop Panda}
\authorrunning{S Panda}
\institute{Northumbria University}
%
\maketitle              
\begin{abstract}
Large Language Models (LLMs), which have gained significant traction in recent years, also function as big structured repositories of data. User personas are a significant and widely utilized method in HCI. This study aims to investigate how LLMs, in their role as data repositories, interpret user personas.  Our focus is specifically on personas within the Indian context, seeking to understand how LLMs would interpret such culturally specific personas. To achieve this, we conduct both quantitative and qualitative analyses. This multifaceted approach allows us a primary understanding of the interpretative capabilities of LLMs concerning personas within the Indian context.

\end{abstract}

\section{Introduction}
User personas are widely used in HCI research to gain a deeper understanding of the user base and improve user-centered design \cite{pruitt2003personas,adlin2006putting,chang2008personas}. These fictional representations of user archetypes are created based on demographic, behavioral, and psychographic characteristics. User personas enable HCI researchers and designers to develop a deeper understanding of the target users. By identifying their needs, desires, and problems these personas facilitate the creation of products and services that better align with user expectations. By using such personas, researchers can analyze user segments, predict their behavior, and tailor their research findings accordingly \cite{miaskiewicz2009preliminary,long2009real}. This approach enhances the relevance, usability, and impact of HCI research, making it more user-centric and applicable in real-world settings.

Large language models (LLMs) have gained immense popularity and achieved remarkable success in recent years \cite{zhao2023survey,hadi2023survey}. These models, such as OpenAI's GPT-4, Meta's Llama or Google's Gemini have been trained on vast amounts of text data from the internet and exhibit the ability to generate coherent and contextually relevant human-like language \cite{radford2019language}. They have gained widespread popularity due to their versatlity and usefulness across numerous applications. They excel (technically) in tasks such as natural language understanding and machine translation and thus they are very effective for chatbots and content generation. Consequently, LLMs have transformed various industries, including journalism, customer service, creative writing among many others.



As these LLMs have the ability to generate human-like text and understand complex contexts \cite{brown2020language}, these models can also aid in creating user personas by analyzing a vast amount of data and then generating accurate and detailed profiles; by simply aligning with the methods by which HCI researchers and designers create user personas. For example, by analyzing online behavior, social media posts, and consumer feedback, LLMs can identify patterns and extract important insights regarding demographics, preferences, and behaviors. This information can then be used to create realistic user personas, representing the characteristics, needs, motivations, and goals of the intended specific user groups. 


Towards this concept of utilizing LLMs to create User Personas, this paper aims to analyze how these LLMs perceive user personas. We specifically look at personas in an Indian context. The contribution of the paper are basically twofold,
\begin{enumerate}
    \item We provide a preliminary quantitative and qualitative analysis of how LLMs' perceive user personas. 
    \item From the analysis, we further provide some insights into how these LLMs' can help designers co-create user personas
\end{enumerate}
\section{Background}
\subsection{On User Personas}
User personas are archetypal representations of target users\cite{pruitt2003personas,chang2008personas}. They are fictonal characters created to represent the different user groups that a designer or a researcher aims to serve. They play a crucial role in user-centric design approaches \cite{miaskiewicz2011personas}, as they enable designers and researchers to better understand and empathize with their target users' needs, goals, and preferences.  These personas  are derived from extensive research \cite{pruitt2003personas,salminen2020literature,jansen2022create,adlin2010essential} such as user interviews, observations, and other user-centered methods, which provide insights into intended users' characteristics, behaviors, and preferences. They capture individuals' demographic details, motivations, goals, and frustrations, allowing designers to empathize with their target users throughout the design process. Several research studies \cite{miaskiewicz2009preliminary,long2009real,salminen2022use,nielsen2013personas} have been explored to understand the effectiveness of user personas in shaping user-centered designs. For instance, in a study \cite{cooper2003face} participants reported improved communication and shared understanding among multidisciplinary design teams when using personas. It highlighted how personas facilitate the alignment of heterogeneous team members around a common user-centric vision. Another study \cite{adlin2006putting} found that personas help designers prioritize design decisions by providing a focal point for design discussions and evaluations.

\subsection{On Large Language Models}
LLMs are a type of neural network-based AI system designed to process and generate human-like text \cite{brown2020language}. These models are trained on massive amounts of data, allowing them to generate responses that are clear, coherent and suitable for the provided context. They often exhibit a brilliant ability to comprehend and generate natural language, enabling them to perform a wide range of tasks including language translation, question-answering, summarization, and even creative writing \cite{radford2019language}. These models have gained significant attention due to their potential to revolutionize various industries and their applications are vast \cite{kasneci2023chatgpt,thirunavukarasu2023large,yang2023fingpt,wang2023exploring,yuan2024chatmusician}, as they can assist language-related tasks that once required peak human involvement. 



\subsection{Why use LLMs for User Personas?}
LLMs are built from large and diverse datasets that include a variety of linguistic and contextual information. Similarly, personas, which are representations of user archetypes, are also developed by gathering data from a variety of sources. Thus, the intersection of these two constructs (LLMs and user personas) presents an intriguing area of study, particularly in understanding how LLMs interpret, perceive and can help develop user personas. Exploring how LLMs interpret user personas can reveal the capabilities and limitations of LLMs in simulating or mimicking human-like interactions and understanding user needs, ultimately improving their use in systems that are personalized and contextually aware \cite{shen2024understanding,hadi2023large}. 



It is important to emphasize that this analysis or its implications does not aim to replace human subjects \cite{dillion2023can}. Instead, it aims to explore how LLMs perceive and can potentially enhance the understanding of user personas and contribute to user-centered design. User personas serve as fictional representations of target users and thus essentially abstractions created by HCI researchers and designers to inform design decisions, rather than reflecting actual individuals \cite{pruitt2003personas,chang2008personas,long2009real}. They condense and generalize user demographic information, behaviors, and motivations into fictional characters. LLMs similarly, possess the capability to process vast amounts of data and uncover patterns and correlations (that humans may overlook) \cite{brown2020language,radford2019language}. They can therefore extract valuable insights from large datasets, contributing to a more comprehensive development of abstract user personas. 



\section{Methods \& Experiments}
To understand how LLM's perceive these user personas, we use both quantitative and qualitative analyses. We choose three India-based user personas from \cite{kaduskar2010understanding}. The personas we choose from \cite{kaduskar2010understanding} include
\begin{enumerate}
    \item \textit{Entertainment Seeker} - Persona A,
    \item \textit{Dependent Family Talker} - Persona B,
    \item \textit{Networker and Information Seeker} - Persona C
\end{enumerate}

For quantitative analysis, we use the existing persona perception scale \cite{salminen2020persona} to how how the llms perceive the personas. The constructs or traits we study include Completeness, Clarity, Consistency and Credibility. We do not use the Wiling To Use construct as we are not specifying any tasks and hence it is not relevant to the task. We also do not use the Empathy, Similarity  or Likability construct as we do not assume the LLM to have any sort of agency that could empathize with or like a persona. As stated earlier, we only understand LLMs as large repositories of data that can respond to certain (mostly technical) constructs of the personas. 

For each of the constructs, we use the corresponding questions, as prescribed in \cite{salminen2020persona}. Some these questions, in Table \ref{tab:example}, are adapted for use for LLMs and for the 3 chosen personas from \cite{kaduskar2010understanding}. 


\begin{table}[h!]
    \centering
    \begin{tabular}{|c|p{8cm}|}
        \hline
        Trait & Questions \\
        \hline
        Completeness & The persona profile is detailed enough to make decisions about the users it describes. \\
         & The persona profile seems complete. \\
         & The persona profile provides enough information to understand the people it describes. \\
         & The persona profile is not missing vital information. \\
        Clarity & The information in the persona profile is easy to understand. \\
        & The information about the persona is well presented. \\
        & The persona is memorable. \\
        Consistency & The persona's demographic information (age,gender,country) corresponds with other information in the persona profile. \\
        & The persona information seems consistent. \\
        Credibility & The persona seems likes a real person. \\
        \hline
    \end{tabular}
    \caption{The questions for each of the traits, borrowed from \cite{salminen2020persona}}
    \label{tab:example}
\end{table}



In our approach to zero-shot prompt design, we adopt a similar strategy as Huang et al. \cite{huang2023chatgpt}. Specifically, we instructed the llms to respond solely with a number corresponding to the levels of the Likert scale. Each level of the Likert scale is clearly delineated and explained to ensure precise and consistent responses. We use two models released from OpenAI\cite{roumeliotis2023chatgpt}, gpt-3.5-turbo-16k and gpt-4 for our quantitative analysis. 

\begin{quote}
\texttt{
    \textbf{Prompt for Quantitative Analysis}:  Say the problem is to understand mobile usage in rural India, and we have to build personas of mobile phone users in India. 
        One of the personas description is , "..."
        How much do you agree with overall description of the persona on a Likert scale of 0 - 7, where 0 is strongly disagree an 7 is strongly agree  [Questions from Table \ref{tab:example}]. Please answer strictly in a numeric format.}
        \\
        
\texttt{    \textbf{Prompt for Qualitative Analysis}: Say the problem is to understand mobile usage in rural, and we have to build personas of mobile phone users in India. 
    Now given this description of the persona, "..." 
    can you predict the demographic profile of the persona across age (a range required), income status, technology competence, occupation and social status?}

\end{quote}

\section{Results \& Analysis}
\subsection{Quantitative Analysis}
We report the mean and standard deviation of the llm outputs after aggregation of the questions across the five chosen constructs. 

\begin{table}[h]
    \centering
\begin{tabular}{ |p{2cm}|p{2.5cm}|p{3cm}|p{2cm}|}
 \hline
 Personas & Characteristic & gpt-3.5-turbo-16k &gpt-4\\
 \hline
 Persona A   
    & Completeness    &6$\pm$0&   6.75$\pm$0.5\\
   & Clarity    &5.5$\pm$0.58&   6.25$\pm$0.96\\
   & Consistency    &6$\pm$0&   7$\pm$0\\
    & Credibility    &5$\pm$0&   6$\pm$0\\ \hline
     Persona B   
    & Completeness    &5.67$\pm$0.58&   6.5$\pm$0.58\\
   & Clarity    &6$\pm$0.82&   6.75$\pm$0.5\\
   & Consistency    &6.5$\pm$0.71&   7$\pm$0\\
    & Credibility    &6$\pm$0&   6$\pm$0\\ \hline
     Persona C   
    & Completeness    &6.25$\pm$0.5&   6.5$\pm$0.58\\
   & Clarity    &5.87$\pm$1.31&   6.75$\pm$0.5\\
   & Consistency    &6.52$\pm$0.68&   7$\pm$0\\
    & Credibility    &6$\pm$0&   7$\pm$0\\
 \hline
\end{tabular}
    \caption{Mean \& Standard deviation of LLMs' responses}
    \label{tab:my_label}
\end{table}

Given the small dataset, we chose not to do any statistical analysis. Rather we chose to look at the aggregated means of LLMs' outputs of the different constructs, 
\begin{enumerate}
    \item Completeness - 5.97, 6.58
    \item Clarity - 5.79, 6.58
    \item Consistency - 6.34, 7
    \item Credibility - 5.67, 6.33
\end{enumerate} 

We can see that both the models score all the personas highest on the Consistency construct. Completeness and clarity constructs also have high scores. Credibility, as expected, scores comparatively low as it may be difficult for the LLM to realize if the persona seems like a real person. Similarly, aggregated means across the different personas are, 
\begin{enumerate}
    \item Persona A - 5.63, 6.5
    \item Persona B - 6.04, 6.56
    \item Persona C - 6.16, 6.81
\end{enumerate}

All personas receive almost consistent scores from both the models with Persona C being the closest.



\subsection{Qualitative Data Analysis}

We employed gpt-4 to recreate the user groups based on the provided scenarios. Towards this we provide the description of the three personas as the prompt and instruct the LLMs to reconstruct the demographic profile of the persona. 

The underlying objective of this approach was to utilize LLMs to determine whether they could accurately reconstruct the demographics of the personas as described. This involved analyzing the extent to which the LLMs could understand and replicate the intricate details of the user groups from the given descriptions, thereby assessing the models' capability to mirror the original demographic characteristics of the personas accurately. 

For each of the personas, we present the text (quoted below) as generated by the LLMs (gpt-4).
\vspace{0.5cm}
\\
For Persona A, 
\begin{quote}

{\fontfamily{qcr}\selectfont\large
{
\textbf{Age:} This persona likely falls into the younger age group, possibly between 15-30 years old. This is due to their interest in games, music, and socializing with friends, which are more common activities among younger individuals.
\\
\\
\textbf{Income Status:} Given their preference for Chinese phones, which are often more affordable, and their lack of close monitoring of phone expenses, this persona may belong to the lower to middle income bracket. They may have enough disposable income to spend on entertainment but not necessarily on high-end devices.
\\
\\
\textbf{Technology Competence:} This persona seems to be fairly tech-savvy. They use their phones for a variety of purposes beyond just making calls, such as playing games and listening to music. They also seem to be aware of the different features that phones offer and choose their phones based on these features.
\\
\\
\textbf{Occupation:} Given their age and usage patterns, this persona could be a student or a young professional. Their occupation likely doesn't require extensive use of the phone for work purposes, as their usage is primarily for entertainment.
\\
\\
\textbf{Social Status:} This persona seems to have a strong social circle and enjoys socializing with friends. They may be considered popular or sociable within their community. They also value their privacy, suggesting a level of maturity and self-awareness.}}
\end{quote}
The age group is estimated to be 15-30 years. Major themes that emerge from the demographic profile include \textit{spending on entertainment}, \textit{students or young professionals},  \textit{playing games, listening to music, enjoys socializing with friends}.  
\vspace{0.5cm}
\\
For Persona B, 
\begin{quote}

{\fontfamily{qcr}\selectfont\large
{
\textbf{Age:} Likely to be older, possibly 50+ years. This is due to their lack of comfort and familiarity with mobile technology, which suggests they did not grow up with it and have not adapted to it as younger generations have.
\\
\\
\textbf{Income Status:} Likely to be low to middle income. The fact that the phone is bought out of necessity and shared among the household suggests that the cost of individual phones for each family member may be prohibitive. 
\\
\\
\textbf{Technology Competence:} Low. The persona is described as being intimidated by the phone and unable to operate it properly, often requiring assistance from others.
\\
\\
\textbf{Occupation:} Likely to be involved in manual labor or agriculture, given the rural setting and the fact that the phone is not used for work purposes. Alternatively, they could be retired or a homemaker.
\\
\\
\textbf{Social Status:} Likely to be lower or middle class, given the income status and occupation. The persona's aspirations for better education for their children suggest they may not have had access to high-quality education themselves, which often correlates with lower social status.}}
\end{quote}
The age group is estimated to be 50+ years. Major themes that emerge from the demographic profile include \textit{low to middle income}, \textit{retired or homemaker or involved in manual labor or agriculture},  \textit{lower social status}.   
\vspace{0.5cm}
\\
For Persona C, 
\begin{quote}

{\fontfamily{qcr}\selectfont\large
{
\textbf{Age:} 25-45 years old. This age range is typically associated with active networking and business activities. 
\\
\\
\textbf{Income Status:} Low to middle income. The persona is concerned about money and keeps a close watch on spending, which suggests a limited income. 
\\
\\
\textbf{Technology Competence:} Moderate to high. The persona uses his mobile phone for networking and business, and solves his own problems, suggesting a good understanding of technology.
\\
\\
\textbf{Occupation:} Likely a small business owner or entrepreneur. The persona uses his phone primarily for business and is a person who others frequently seek help from, suggesting a leadership role or a position of influence.
\\
\\
\textbf{Social Status:} Middle class. The persona's concern about money and the fact that he plans and executes his spending suggests that he is not wealthy. However, his active networking and the fact that people look up to him or seek help from him suggest a certain level of social standing.}}
\end{quote}
The age group is 25-45 years. Major themes that emerge from this demographic profile are \textit{limited income}, \textit{good understanding of tech and a small business owner or entrepreneur} and \textit{middle class status}.  


\subsection{Co-creation of personas with LLMs}
Given the above findings, we present here some actionable insights of how LLMs can be used to help create User Personas. 

\begin{enumerate}
    \item LLMs possess the capability to \textbf{effectively segregate user groups} by focusing on releavnt user populations. By analyzing vast amounts of data, LLMs can identify distinct user characteristics, preferences, and behavior patterns, thereby enabling the creation of accurate user segments. 
    \item LLMs can serve as robust tools for \textbf{evaluating, validating, or verifying descriptions} of such personas. For instance, to validate a persona, an LLM can be tasked with \textit{impersonating an individual} of the described user group and subsequently performing various tasks or responding to questions. This process tests the coherence and accuracy of the persona's characteristics as delineated in the description. By simulating real-life scenarios and interactions, LLMs can also help identify inconsistencies, gaps, or inaccuracies in the personas, thus providing an extensive validation mechanism. 
    \item LLMs can significantly \textbf{enhance existing descriptions} of personas by \textit{continuing to write} thus introducing variety and detail. These augmentations are not to be incorporated directly into the descriptions, but they serve as valuable cues for further refinement and elaboration of the persona descriptions. By leveraging the extensive linguistic and contextual knowledge embedded in LLMs, additional layers of details, nuance and ideas can be added to personas, making them more comprehensive and data-driven.
\end{enumerate}
\section{Discussion}

Our research contribution aligns with \cite{salminen2020persona} and \cite{huang2023chatgpt}. We try out the different constructs/traits of the persona perception scale on India-based user personas \cite{kaduskar2010understanding} and use two LLM models from OpenAI. The results from both our quantitative and qualitative studies indicate that both the LLMs are generally in agreement with the three personas. This is supported by the scores observed in the quantitative analysis and the coherent, consistent themes that emerged from the qualitative analysis.

\subsection{Limitations}

With the small dataset, it is difficult to find any statistically significant differences between the different personas or among the traits themselves. Nonetheless, these results are suggestive and warrant further analysis. The output of the LLMs an vary depending on the hyper-parameters and each iteration. While experimenting with different hyperparameter settings can lead to more consistent outputs, replicating results exactly remains challenging. Consequently, LLMs are most effectively utilized for indicating or approximating results rather than providing exact replications. This aspect should also be taken into account while interpreting the results. Further, there are biases observed in LLMs \cite{navigli2023biases,liang2021towards,kotek2023gender} are often manifested through various dimensions, including but not limited to gender, race, ethnicity, religion, and socio-economic status.  These can also extend beyond explicit categories to encompass subtler forms of bias, such as linguistic biases and cultural biases.

\subsection{Future Work}


Future work will use the collection of a more extensive set of user personas, with a particular emphasis on those situated within the Indian context. This initiative aims to diversify the dataset, thereby enhancing the cultural and contextual relevance of the personas. By incorporating a broader range of user experiences and characteristics from India, we can ensure that the models are better equipped to handle the unique social, economic, and cultural dynamics present in the country. In addition, future work will involve the refinement of prompting strategies and the exploration of innovative methods for engaging different LLM models with personas. This will include developing more sophisticated techniques for eliciting detailed and accurate responses from the models, thereby improving the precision and depth of the personas. Also, new approaches will be investigated to tailor these personas more effectively to specific design needs, ensuring that they are not only accurate but also highly relevant to the contexts in which they are applied. 



\section{Conclusion}

In this study, we set out to explore how LLMs understand and interpret personas. We focused on three personas from an Indian context to see how these models would respond. For the quantitative analysis  we used a persona perception scale to measure how different LLM models viewed these personas. For the qualitative analysis we looked at how accurately the LLMs could recreate the personas' demographics given the scenarios. We used zero-shot prompting strategy for both the analysis. We found that the LLMs had scored the personas highly in the perception scale and were able to almost correctly replicate the demographics of the personas. We also shared some practical insights on how LLMs can be used to develop user personas.


\section*{Ethics Statement}
Parts of text {\fontfamily{qcr}\selectfont\large quoted} in section 4.2 have been generated by AI. This was integral to the paper, as the text generation itself is a part of the research contribution. 

\bibliographystyle{splncs04}
\bibliography{paper}
%




\end{document}